# Probing the mass composition of TeV cosmic rays with HAWC

**J.C. Arteaga-Velázquez[a,*] for the HAWC Collaboration**

[a] *Instituto de Física y Matemáticas,*
*Universidad Michoacana, Morelia, Michoacan, Mexico*

*E-mail:* juan.arteaga@umich.mx

In this contribution, we have investigated the energy spectra of the elemental mass groups of cosmic rays for the Light (H+He), medium (C+O) and heavy (Ne-Fe) components using the High Altitude Water Cherenkov Gamma-Ray observatory (HAWC). The study was carried out in the energy interval from 10 TeV to 1 PeV using almost 5 years of data on hadronic air showers. The energy spectra were unfolded using the bidimensional distribution of the lateral shower age versus the reconstructed primary energy. We have employed the QGSJET-II-04 high-energy hadronic interaction model for the current analysis. The results show the presence of fine structure in the spectra of the light, medium and heavy mass groups of cosmic rays.

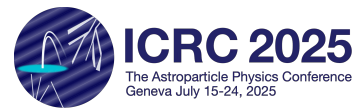

*Speaker

## 1. Introduction

The study of the composition, energy spectrum and arrival directions of TeV cosmic rays are of importance to investigate the high-energy phenomena that occur in our galaxy. To this end, direct and indirect detection techniques of cosmic rays can be conveniently applied. HAWC is a high-altitude air-shower observatory with capabilities to study cosmic rays between a few TeV up to 1 PeV [1]. It has allowed the measurement of the all-particle energy spectrum of cosmic rays between 10 TeV and 1 PeV [2, 3], the elemental spectrum of H+He nuclei from 6 TeV to 158 TeV [4], and the large- and small-scale anisotropies in the cosmic-ray skymaps in the 3-73 TeV interval [5–7]. Recently, HAWC capabilities have been applied to separate the elemental mass groups of cosmic rays. In a preliminary study, HAWC determined the energy spectra for the H, He and heavy (C-Fe) primaries of cosmic rays at TeV energies [8]. The results showed fine structure in these spectra, specifically, softenings at around tens (hundreds) of TeV for the light (heavy) nuclei. In addition, it also detected individual hardenings close to 100 TeV in the spectra of H and He primaries. In order to continue the investigations, here, we have carried out a preliminary analysis to determine the spectra of the light (H+He), medium (C+O) and heavy (Ne-Fe) mass groups.

## 2. The HAWC observatory and the experimental data

HAWC is located on a plateau between the Pico de Orizaba and the Sierra Negra volcanos (19° N, 97.3° W) in Puebla, Mexico at an altitude of 4100 m (637 g/cm$^2$ of vertical atmospheric depth) [1]. The primary detector consists of 300 water Cherenkov detectors (WCD) with a diameter of 7.3 m and a height of 5.4 m installed over a surface area of 22000 m$^2$. Each WCD contains 4 photomultipliers (PMT) and 200, 000 $\ell$ of water. HAWC measures the arrival times of the shower particles at the WCD and the effective charge collected by the PMTs due to the air-shower passage. The arrival direction of the event is determined from the recorded timing of the particles of the extensive air shower (EAS), while the shower core is obtained from the effective charge spatial distribution [9]. These parameters are used to reconstruct the lateral distribution (LDF) of the effective charge, which is estimated as a function of the radial distance from the core in the shower plane. The amplitude and the slope (lateral shower age) of the LDF are calculated from a $\chi^2$ fit to the radial distribution of the effective charge with a NKG-like function [10]. The primary energy of a hadronic showers is derived by means of a maximum likelihood estimation which is applied even-by-event on the measured LDFs [2]. For the energy assignment, the LDF data is compared with MC templates built with proton simulations and the QGSJET-II-04 hadronic interaction model [11].

The full primary array of HAWC has continuously taken data since March 20, 2015. The present analysis is based on a set of $4.62 \times 10^{10}$ selected events taken during the observation period from January 2016 to January 2021 during an effective time of $\Delta t = 1727$ days. These data passed several selection criteria to increase the precision of the study. In particular, only events with a successful reconstruction were considered, with reconstructed energies $3.6 \leq \log_{10}(E_{rec}/\text{GeV}) \leq 6.2$. Events of PeV energies, which are misreconstructed, were removed from the data set with a cut on the logarithm of the LDF amplitude $\log_{10} A < 2.5$. In addition, showers with zenith angles $\theta$ from 20° to 45° were kept. Vertical events were discarded to avoid large systematic errors in the results at energies close to 1 PeV. Showers measured during data taking periods with more than 900 available PMTs and that activated a fraction $f_{hit} \geq 0.2$ of the available channels were also considered. Migration effects due to the uncertainty in the location of the EAS cores were reduced by removing events with shower ages ($s$) greater than $-1.99 f_{hit} + 4$ and with less than 40 hit PMTs within a radius of 40 m from the shower core. Finally, only data with shower ages greater or equal to 1.0 were included.

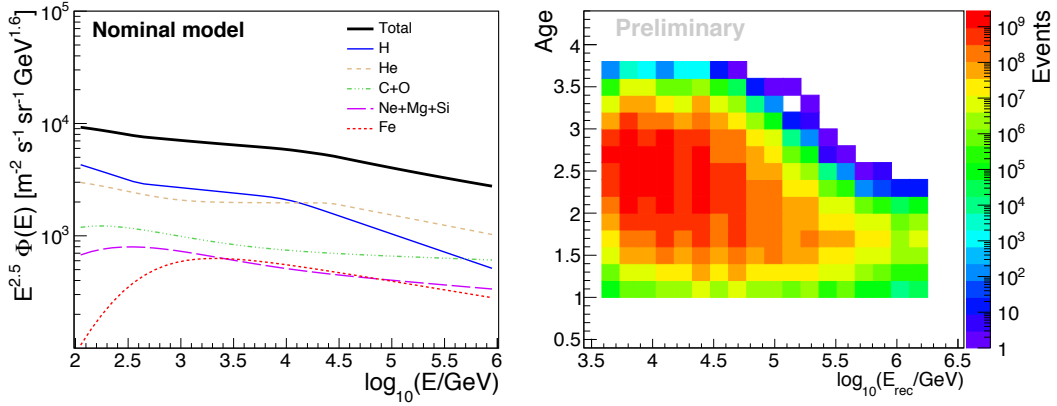

**Figure 1:** Left: Energy spectra of the nominal cosmic-ray composition model used in the present analysis. Right: The shower age vs reconstructed energy distribution for the measured data. Bin sizes of $\Delta s = 0.2$ and $\Delta \log_{10}(E_{rec}/\text{GeV}) = 0.15$ were employed to build the distribution. Selection cuts have been applied as described in the text.

## 3. MC simulations

For the present analysis, MC simulations were generated, which include details about the production and propagation in the atmosphere of EAS, calculated with CORSIKA v7.40 [12], as well as the interactions of the air showers with the detector components, using the Geant4 code [13]. The output is provided in the same format as the measured data and is reconstructed with the algorithms that were applied on the experimental events. The EAS were simulated using the hadronic interaction models FLUKA (hadronic energies $E_h < 80\,\text{GeV}$) and QGSJET-II-04 ($E_h \geq 80\,\text{GeV}$) without thinning for the HAWC site including the effects of the geomagnetic field. Simulations were carried out for zenith angles smaller than 65° and eight primary nuclei, in particular, H, He, C, O, Ne, Mg, Si and Fe, at energies in the 5 GeV – 2 PeV interval. The MC data were produced with an $E^{-2}$ energy spectrum and were reweighted to simulate a nominal model of cosmic-ray composition that was calibrated with direct and indirect measurements (see Fig. 1, left). The data employed to build the model were fitted with broken power-law functions and were measured by the PAMELA [14, 15], AMS-2 [16–19], ATIC-2 [20], CREAM I-II [21, 22], CALET [23–25], DAMPE [26, 27], NUCLEON [28] and KASCADE [29] cosmic-ray experiments.

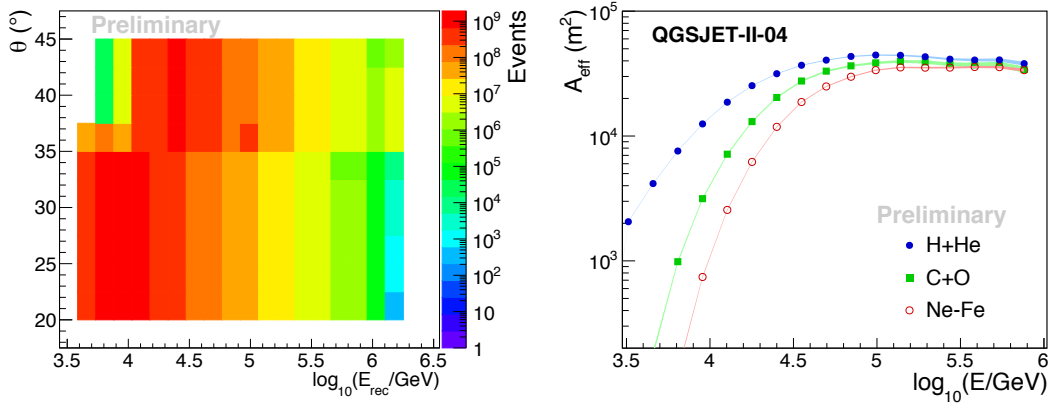

**Figure 2:** Left: The experimental distribution for $\theta$ vs $\log_{10} E_{rec}$. Right: The effective areas for the H+He, C+O and Ne-Fe mass groups of cosmic rays as a function of $\log_{10} E$ estimated with MC simulations using the nominal composition model.

## 4. Analysis procedure

To extract the energy distributions of the H+He, C+O and Ne-Fe mass groups from the selected data, we applied the Gold unfolding method [29, 30] to the 2D-distribution of the shower age versus the estimated primary energy $n(s, \log_{10} E_{rec})$ built for the measured data. This distribution is shown in Fig. 1, right. The objective of the unfolding procedure is to solve the following set of coupled equations:

$$n(s, \log_{10} E_{rec}) = \sum_j \sum_E P_j(s, \log_{10} E_{rec} | \log_{10} E)\, n_j(E), \qquad (1)$$

for $n_j(E)$ ($j = 1, 2, 3$), which represents the unfolded energy distribution for each cosmic-ray mass group. Here, $P_j(s, \log_{10} E_{rec} | \log_{10} E)$ is the response matrix for the $j$th group of primaries, which provides the probability that a cosmic ray of true energy $\log_{10} E$ from this mass group generates an EAS with shower age $s$ and reconstructed energy $\log_{10} E_{rec}$, which can trigger and pass the selection cuts of HAWC. The response matrices were computed using the QGSJET-II-04 MC simulations for the nominal cosmic-ray composition model.

Eq. (1) was solved iteratively with the Gold unfolding algorithm. The initial guess distributions for $n_j(E)$ were taken from the nominal model. As regularization procedure, the intermediate energy distributions were smoothed with broken power-law functions. To evaluate the quality of the unfolded distributions, we estimated the $\chi^2$ between the measured and the estimated $s$ vs $\log_{10} E_{rec}$ distributions using MINUIT [31] from ROOT [32]. The $\chi^2$ slowly decreases with the number of iterations, but no minimum is observed. Therefore, we added more information to the analysis to find the proper iteration depth. In particular, we employed the 2D-distribution for the zenith angle $\theta$ vs the estimated energy $\log_{10} E_{rec}$ (see Fig. 2, left), which also has mass information. Calculating the $\chi^2$ between the experimental data and the unfolded solutions, we found a minimum for 23 iterations. Once the $n_j(E)$ distributions were obtained, the corresponding energy spectra $\Phi_j(E)$ were calculated by means of the formula $\Phi_j(E) = n_j(E)/[\Delta E\, \Delta\Omega\, \Delta t\, A_{\text{eff},j}(E)]$, where $\Delta\Omega = 1.46$ sr is the solid angle range, $\Delta E$ is the width of the bin centered at the energy $E$, and $A_{\text{eff},j}(E)$ is the effective area for the $j$ primary group of cosmic ray (c.f. Fig. 2, left), which was estimated with the MC simulations.

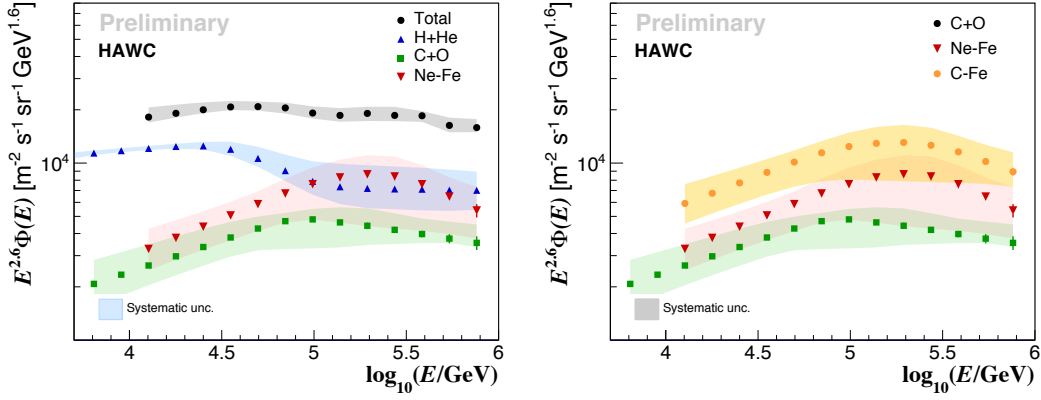

**Figure 3:** Left: The total, H+He, C+O and Ne-Fe energy spectra of cosmic rays, which were obtained in this analysis are shown with black circles, blue upward triangles, green squares and red downward triangles, respectively. Right: The C+O, Ne-Fe and heavy (C-Fe) spectra of cosmic rays are plotted with green squares, red downward triangles and orange circles, correspondingly. Statistical errors are displayed with vertical error bars, and systematic errors, with error bands.

## 5. Results and discussions

Fig. 3, left, shows the unfolded energy spectra for H+He, C+O and Ne-Fe nuclei, along with the total spectrum obtained from the sum of these mass groups. Meanwhile, Fig. 3, right, displays the result for the heavy component (C-Fe) of cosmic rays, which is the sum of the C+O and Ne-Fe spectra, also shown in the plot. The spectra are presented with total statistical and systematic errors. The total statistical uncertainties are not larger than 10% and include the errors due to the experimental statistics and the limited size of the MC sample added in quadrature. Regarding the total systematic errors, they are smaller than 50% and have contributions from uncertainties in the modeling of the PMTs and the HAWC configuration [10], the composition of cosmic rays, the unfolding method and the effective areas. The error due to the uncertainties in the cosmic-ray composition was calculated by replacing the nominal model by alternative ones to evaluate the response matrices and effective areas in the procedure. In particular, we employed the GSF [33] and H3a [34] models, as well as spectra based on AMS-II [16–19] and ATIC-2 [20] data. To estimate the systematic errors from the unfolding method, the results of the Gold unfolding procedure were compared with those from the Bayesian unfolding algorithm [35], and also the results obtained with the priors from the nominal model were compared with those estimated with the different composition models. Finally, the error due to the effective areas were calculated by varying $A_{\text{eff}}$ for each mass group within their corresponding uncertainties and then by repeating the calculations of the energy spectra. All these contributions were added in quadrature to estimate the total systematic error.

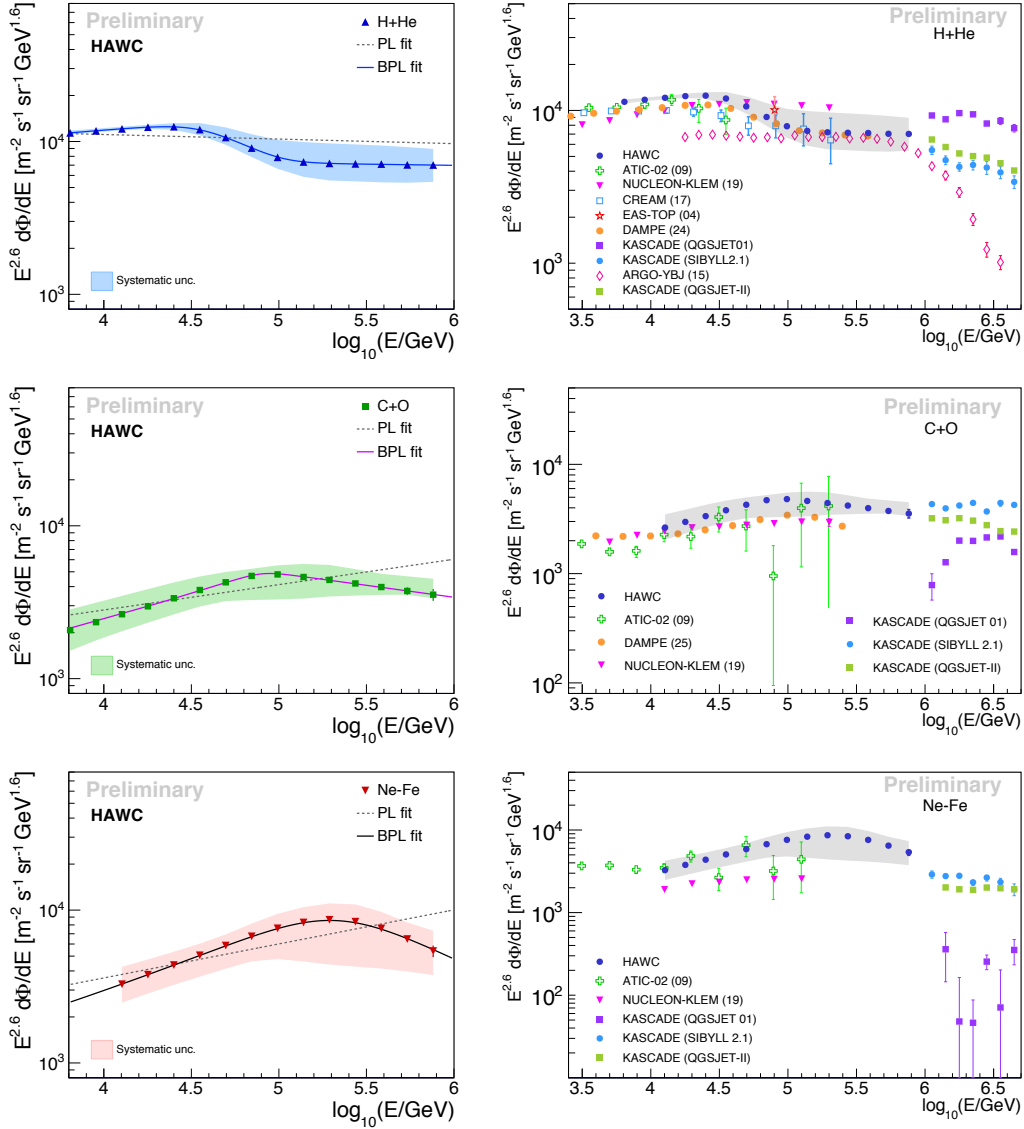

**Figure 4:** Left: Results of the fits with the broken power law (BPL) of eq. (2) and a power law (PL) to the HAWC energy spectra of Fig. 3. Right: Comparison of the HAWC energy spectra with data from direct [20, 28, 36–38] and indirect [29, 39–41] cosmic-ray experiments. In both columns, the error bands represent HAWC systematic uncertainties. From top to bottom, each row display the results for H+He, C+O and Ne-Fe spectra.

The energy spectra of the H+He, C+O and N-Fe mass groups of cosmic rays do not follow a simple power-law behavior, as seen from Fig. 3, due to the presence of different breaks. The spectrum for the light (H+He) component exhibits a softening between 10 and 100 TeV and a hardening close to 100 TeV, while the spectra for the C+O, Ne-Fe and heavy (C-Fe) mass groups show individual cutoffs in the 100 TeV − 1 PeV interval. To characterize these features, the energy spectra were fitted with a broken power-law function

$$\Phi(E) = \Phi_0 \, E^{\gamma_0} \left[1 + \left(\frac{E}{E_0}\right)^{\varepsilon_0}\right]^{(\gamma_1-\gamma_0)/\varepsilon_0} \left[1 + \left(\frac{E}{E_1}\right)^{\varepsilon_1}\right]^{(\gamma_2-\gamma_1)/\varepsilon_1}, \qquad (2)$$

using the $\chi^2$ method and taking into consideration the covariance among the data points that appears from the unfolding method. The covariance matrices were estimated with the Bootstrap method [42]. In Eq. 2, $\Phi_0$, $\gamma_j$ ($j = 1, 2, 3$), $E_i$ and $\varepsilon_i$ ($i = 0, 1$) are free parameters. $\Phi_0$ is the normalization factor, $E_0$ ($E_1$), the location of the break in the $10-10^2$ TeV ($10^2-10^3$ TeV) decade and $\varepsilon_0$ ($\varepsilon_1$), the smoothing parameter of the break at $E_0$ ($E_1$). The spectral indexes before and after the first (second) break are $\gamma_0$ ($\gamma_1$) and $\gamma_1$ ($\gamma_2$), respectively.

The results of the fits (see Fig. 4, left column) reveal that the softening of the light mass group is located at $E_0 = (31.6 \pm 0.2)$ TeV and is produced by a change in the spectral index of $\Delta\gamma_{10} = \gamma_1 - \gamma_0 = -0.57 \pm 0.01$. The corresponding hardening was found at $E_1 = (110.1^{+3.8}_{-3.7})$ TeV. Here the spectral index change from $\gamma_1 = -3.08 \pm 0.01$ to $\gamma_2 = -2.65 \pm 0.02$. The fits also show that the cut-off in the C+O spectrum appears at energies $E_1 = (81.0 \pm 1.7)$ TeV with $\Delta\gamma_{21} = \gamma_2 - \gamma_1 = -0.51 \pm 0.01$. For the Ne-Fe component, the cut-off was located at higher energies ($E_1 = 210^{+6}_{-5}$ TeV). Here $\Delta\gamma_{21} = \gamma_2 - \gamma_1 = -1.07 \pm 0.07$.

Fig. 3 supports the results on the features of the H+He and heavy (C-Fe) spectra previously reported in [4] and [8]. But also shows that the break in the heavy component is produced by the superposition of the cut-offs in the C+O and Ne-Fe spectra. The latter seems to be more abundant than the C+O component at hundred cosmic rays. Here, it can be also seen that the TeV cut-offs for the heavier mass groups are located at higher energies. Finally, in Fig. 4, right column, HAWC results are compared with the data from direct and indirect cosmic-ray measurements. At TeV energies HAWC H+He spectrum is in agreement with ATIC-2 [20]. Above $\sim 30$ TeV, it agrees with CREAM [36], DAMPE [37], EAS-Top [39] and ARGO-YBJ [40]. The C+O (Ne-Fe) are in agreement with ATIC-02 [20] between 10 and 130 (200) TeV, but above NUCLEON measurements [28].

## 6. Conclusions

With HAWC data on the lateral shower age and the primary energy of hadronic EAS events and using unfolding techniques, the energy spectra for the H+He, C+O and Ne-Fe mass groups of cosmic rays were estimated in two decades of energy, from 10 TeV to 1 PeV. The results show that the energy spectra of these primary groups have different features. The energy spectrum for the H+He mass group shows a softening at $(31.6 \pm 0.2)$ TeV and a hardening close to $(110.1^{+3.8}_{-3.7})$ TeV, while the C+O and Ne-Fe spectra show individual cut-offs at $(81.0 \pm 1.7)$ TeV and $(210^{+6}_{-5})$ TeV, respectively. The superposition of the cut-offs in the C+O and Ne-Fe intensities produce a knee-like feature in the spectra of the heavy (C-Fe) component of cosmic rays.

**Acknowledgments**. We acknowledge the support from: the US National Science Foundation (NSF); the US Department of Energy Office of High-Energy Physics; the Laboratory Directed Research and Development (LDRD) program of Los Alamos National Laboratory; Consejo Nacional de Ciencia y Tecnología (CONACyT), México, grants LNC-2023-117, 271051, 232656, 260378, 179588, 254964, 258865, 243290, 132197, A1-S-46288, A1-S-22784, CF-2023-I-645, CBF2023-2024-1630, cátedras 873, 1563, 341,

323, Red HAWC, México; DGAPA-UNAM grants IG101323, IN111716-3, IN111419, IA102019, IN106521, IN114924, IN110521, IN102223; VIEP-BUAP; PIFI 2012, 2013, PROFOCIE 2014, 2015; the University of Wisconsin Alumni Research Foundation; the Institute of Geophysics, Planetary Physics, and Signatures at Los Alamos National Laboratory; Polish Science Centre grant, 2024/53/B/ST9/02671; Coordinación de la Investigación Científica de la Universidad Michoacana; Royal Society - Newton Advanced Fellowship 180385; Gobierno de España and European Union-NextGenerationEU, grant CNS2023- 144099; The Program Management Unit for Human Resources & Institutional Development, Research and Innovation, NXPO (grant number B16F630069); Coordinación General Académica e Innovación (CGAI-UdeG), PRODEP-SEP UDG-CA-499; Institute of Cosmic Ray Research (ICRR), University of Tokyo. H.F. acknowledges support by NASA under award number 80GSFC21M0002. C.R. acknowledges support from National Research Foundation of Korea (RS-2023-00280210). We also acknowledge the significant contributions over many years of Stefan Westerhoff, Gaurang Yodh and Arnulfo Zepeda Domínguez, all deceased members of the HAWC collaboration. Thanks to Scott Delay, Luciano Díaz and Eduardo Murrieta for technical support.

## References

[1] A.U. Abeysekara et al., HAWC Collab., NIM A1052, (2023) 168253.
[2] R. Alfaro et al., HAWC Collab., Phys. Rev. D 96 (2017) 122001.
[3] A. Albert et al., HAWC Collab., Astrop. Phys. 167 (2025) 103077.
[4] A. Albert et al., HAWC Collab., PRD 105 (2022), 063021.
[5] A.U. Abeysekara et al., HAWC Collab., Astrophys. J. 796 (2014), 108.
[6] A.U. Abeysekara et al., HAWC Collab., Astrophys. J. 865 (2018), 57.
[7] A.U. Abeysekara et al., HAWC Collab., Astrophys. J. 871 (2019), 096.
[8] J.C. Arteaga-Velázquez et al., HAWC Collab., PoSICRC2023(2023)299.
[9] A.U. Abeysekara et al., HAWC Collab., Astrophys. J. 843 (2017), 39.
[10] A. U. Abeysekara et al., Astrophys. J. 881 (2019) 134.
[11] S. Ostapchenko, Phys. Rev. D 83 (2011) 014018.
[12] D. Heck et al., CORSIKA: A Monte Carlo Code to Simulate Extensive Air Showers, FZK Berichte 6019, Karlsruhe, Germany, 1998.
[13] S. Agostinelli et al., NIMA 506 (2003) 250.
[14] O. Adriani et al., PAMELA Collab., Science 332 (2011) 69.
[15] O. Adriani et al., PAMELA Collab., ApJ 791 (2014) 93.
[16] M. Aguilar et al., AMS Collab., Phys. Rev. Lett. 115 (2015) 211101.
[17] M. Aguilar et al., AMS Collab., Phys. Rev. Lett. 119 (2017) 251101.
[18] M. Aguilar et al., AMS Collab., Phys. Rev. Lett. 124 (2020) 211102.
[19] M. Aguilar et al., AMS Collab., Phys. Rev. Lett. 126 (2021) 041104.
[20] A. D. Panov et al., ATIC-2 Collab., Bull. Russ. Acad. Sci. Phys. 73, No. 5 (2009) 564.
[21] H. S. Ahn et al., CREAM Collaboration, Astrophys. J. 707 (2009) 593.
[22] Y. S. Yoon et al., CREAM Collaboration, Astrophys. J. 728 (2011) 122.
[23] O. Adriani et al., CALET Collab., Phys. Rev. Lett. 122 (2019) 181102.
[24] O. Adriani et al., (CALET Collaboration), Phys. Rev. Lett. 126 (2021) 241101.
[25] P. Brogi et al., CALET Collab., PoS ICRC2021 (2021) 101.
[26] Q. An et al., DAMPE Collab., Science Advances 5, No. 9 (2019) eaax3793.
[27] F. Alemanno et al., DAMPE Collab., PRL 126 (2021) 201102.
[28] E. V. Atkin et al., NUCLEON Collaboration, Astron. Rep. 63 (2019) 66.
[29] T. Antoni et al., KASCADE Collab., Astrop. Phys. 24 (2005) 1.
[30] R. Gold, An iterative unfoding method for response matrices, Report ANL-6984, Argonne National Laboratory, USA,1964.
[31] F. James, MINUIT - Function Minimization and Error Analysis Reference Manual, CERN Report number: CERN-D506.
[32] R. Brun and F. Rademakers, NIMA 389 (1997) 81.
[33] H. P. Dembinski et al., PoS(ICRC2017) 533.
[34] T. K. Gaisser, Astropart. Phys. 35 (2012) 801.
[35] G. D'Agostini, A multidimensional unfolding method based on Bayes' theorem, NIMA 362, (1995) 487.
[36] Y. S. Yoon et al., CREAM Collaboration, Astrophys. J. 839 (2017) 5.
[37] F. Alemanno et al., DAMPE Collaboration, Phys. Rev. D 109 (2024) L121101.
[38] I. Cagnoli et al., DAMPE Collaboration, EPJ Web of Conf. 319, (2025) 02002.
[39] A. Aglietta et al., EAS-Top Collaboration, Astropart. Phys. 21 (2004) 223.
[40] B. Bartoli et al., ARGO-YBJ Collaboration, PRD 92 (2015) 092005.
[41] W.D. Apel et al., KASCADE-Grande Collaboration, Astropart. Phys. 47 (2013) 54.
[42] J.A. Rice, Mathematical Statistics and Data Analysis, 3rd ed. (Thomson Brooks/Cole, Belmont, 2010).

## Full Author List: HAWC Collaboration

R. Alfaro[1], C. Alvarez[2], A. Andrés[3], E. Anita-Rangel[3], M. Araya[4], J.C. Arteaga-Velázquez[5], D. Avila Rojas[3], H.A. Ayala Solares[6], R. Babu[7], P. Bangale[8], E. Belmont-Moreno[1], A. Bernal[3], K.S. Caballero-Mora[2], T. Capistrán[9], A. Carramiñana[10], F. Carreón[3], S. Casanova[11], S. Coutiño de León[12], E. De la Fuente[13], D. Depaoli[14], P. Desiati[12], N. Di Lalla[15], R. Diaz Hernandez[10], B.L. Dingus[16], M.A. DuVernois[12], J.C. Díaz-Vélez[12], K. Engel[17], T. Ergin[7], C. Espinoza[1], K. Fang[12], N. Fraija[3], S. Fraija[3], J.A. García-González[18], F. Garfias[3], N. Ghosh[19], A. Gonzalez Muñoz[1], M.M. González[3], J.A. Goodman[17], S. Groetsch[19], J. Gyeong[20], J.P. Harding[16], S. Hernández-Cadena[21], I. Herzog[7], D. Huang[17], P. Hüntemeyer[19], A. Iriarte[3], S. Kaufmann[22], D. Kieda[23], K. Leavitt[19], H. León Vargas[1], J.T. Linnemann[7], A.L. Longinotti[3], G. Luis-Raya[22], K. Malone[16], O. Martinez[24], J. Martínez-Castro[25], H. Martínez-Huerta[30], J.A. Matthews[26], P. Miranda-Romagnoli[27], P.E. Mirón-Enriquez[3], J.A. Montes[3], J.A. Morales-Soto[5], M. Mostafá[8], M. Najafi[19], L. Nellen[28], M.U. Nisa[7], N. Omodei[15], E. Ponce[24], Y. Pérez Araujo[1], E.G. Pérez-Pérez[22], Q. Remy[14], C.D. Rho[20], D. Rosa-González[10], M. Roth[16], H. Salazar[24], D. Salazar-Gallegos[7], A. Sandoval[1], M. Schneider[1], G. Schwefer[14], J. Serna-Franco[1], A.J. Smith[17], Y. Son[29], R.W. Springer[23], O. Tibolla[22], K. Tollefson[7], I. Torres[10], R. Torres-Escobedo[21], R. Turner[19], E. Varela[24], L. Villaseñor[24], X. Wang[19], Z. Wang[17], I.J. Watson[29], H. Wu[12], S. Yu[6], S. Yun-Cárcamo[17], H. Zhou[21],

[1]Instituto de Física, Universidad Nacional Autónoma de México, Ciudad de Mexico, Mexico, [2]Universidad Autónoma de Chiapas, Tuxtla Gutiérrez, Chiapas, México, [3]Instituto de Astronomía, Universidad Nacional Autónoma de México, Ciudad de Mexico, Mexico, [4]Universidad de Costa Rica, San José 2060, Costa Rica, [5]Universidad Michoacana de San Nicolás de Hidalgo, Morelia, Mexico, [6]Department of Physics, Pennsylvania State University, University Park, PA, USA, [7]Department of Physics and Astronomy, Michigan State University, East Lansing, MI, USA, [8]Temple University, Department of Physics, 1925 N. 12th Street, Philadelphia, PA 19122, USA, [9]Universita degli Studi di Torino, I-10125 Torino, Italy, [10]Instituto Nacional de Astrofísica, Óptica y Electrónica, Puebla, Mexico, [11]Institute of Nuclear Physics Polish Academy of Sciences, PL-31342 11, Krakow, Poland, [12]Dept. of Physics and Wisconsin IceCube Particle Astrophysics Center, University of Wisconsin—Madison, Madison, WI, USA, [13]Departamento de Física, Centro Universitario de Ciencias Exactase Ingenierias, Universidad de Guadalajara, Guadalajara, Mexico, [14]Max-Planck Institute for Nuclear Physics, 69117 Heidelberg, Germany, [15]Department of Physics, Stanford University: Stanford, CA 94305–4060, USA, [16]Los Alamos National Laboratory, Los Alamos, NM, USA, [17]Department of Physics, University of Maryland, College Park, MD, USA, [18]Tecnologico de Monterrey, Escuela de Ingeniería y Ciencias, Ave. Eugenio Garza Sada 2501, Monterrey, N.L., Mexico, 64849, [19]Department of Physics, Michigan Technological University, Houghton, MI, USA, [20]Department of Physics, Sungkyunkwan University, Suwon 16419, South Korea, [21]Tsung-Dao Lee Institute & School of Physics and Astronomy, Shanghai Jiao Tong University, 800 Dongchuan Rd, Shanghai, SH 200240, China, [22]Universidad Politecnica de Pachuca, Pachuca, Hgo, Mexico, [23]Department of Physics and Astronomy, University of Utah, Salt Lake City, UT, USA, [24]Facultad de Ciencias Físico Matemáticas, Benemérita Universidad Autónoma de Puebla, Puebla, Mexico, [25]Centro de Investigación en Computación, Instituto Politécnico Nacional, México City, México, [26]Dept of Physics and Astronomy, University of New Mexico, Albuquerque, NM, USA, [27]Universidad Autónoma del Estado de Hidalgo, Pachuca, Mexico, [28]Instituto de Ciencias Nucleares, Universidad Nacional Autónoma de Mexico, Ciudad de Mexico, Mexico, [29]University of Seoul, Seoul, Rep. of Korea, [30]Departamento de Física y Matemáticas, Universidad de Monterrey, Av. Morones Prieto 4500, 66238, San Pedro Garza García NL, Mexico